\documentclass[review,5p,times,twocolumn,sort&compress]{elsarticle}

\usepackage{amssymb}
\usepackage[utf8]{inputenc}
\usepackage[usenames]{color}
\usepackage{mathptmx}
\usepackage{array}
\usepackage{amsmath}
\usepackage{graphicx}
\usepackage[colorlinks=true] {hyperref}
\usepackage{xspace}
\usepackage{tabularx}
\usepackage{color}
\usepackage[TABBOTCAP]{subfigure}
\usepackage{float}
\usepackage{multirow}
\usepackage{colortbl}
\usepackage{bm}

\journal{Superlattices and Microstructures}

\newcommand{\LMOPS}{Laboratoire Matériaux Optiques, Photonique et Systèmes}
\newcommand{\LMOPSaddress}{Metz, F-57070, France}

\newcommand{\LMOPSUL}{Universit\'e de Lorraine, \LMOPS, \LMOPSaddress}

\newcommand{\LMOPSCS}{\LMOPS, CentraleSupélec, Université Paris-Saclay, \LMOPSaddress}

\newcommand{\E}[2]{#1\times10^{#2}}
\newcommand{\Unit}[2]{ {#1}^{(\mathrm{ #2})}  }

\newcommand{\Atlas}{Atlas$^{\text{\textregistered}}$\xspace}
\newcommand{\Silvaco}{Silvaco$^{\text{\textregistered}}$\xspace}

\newcommand{\X}{{x}}
\newcommand{\Ln}{L_n}
\newcommand{\Lp}{L_p}
\newcommand{\Li}{L_i}
\newcommand{\WF}{W_f}
\newcommand{\Nd}{N_d}
\newcommand{\Na}{N_a}
\newcommand{\Ni}{N_i}

\newcommand{\micron}{\mathrm{\mu m}}

\begin{document}

\begin{frontmatter}



\title{Simulation study of a new InGaN p-layer free Schottky Based Solar Cell}

\author[afful,affcs]{Abdoulwahab Adaine}
\ead{abdoulwahab.adaine@univ-lorraine.fr}

\author[afful,affcs]{Sidi Ould Saad Hamady}
\ead{sidi.hamady@univ-lorraine.fr}

\author[afful,affcs]{Nicolas Fressengeas\corref{cor1}}
\ead{nicolas@fressengeas.net}

\cortext[cor1]{Corresponding author}

\address[afful]{\LMOPSUL}

\address[affcs]{\LMOPSCS}

\begin{abstract}

On the road towards next generation high efficiency solar cells, the ternary Indium Gallium Nitride (InGaN) alloy is a good passenger since it allows to cover the whole solar spectrum through the change in its Indium composition. The choice of the main structure of the InGaN solar cell is however crucial. Obtaining a high efficiency requires to improve the light absorption and the photogenerated carriers collection that depend on the layers parameters, including the Indium composition, p- and n-doping, device geometry\dots Unfortunately, one of the main drawbacks of InGaN is linked to its p-type doping, which is very difficult to realize since it involves complex technological processes that are difficult to master and that highly impact the layer quality.

In this paper, the InGaN p-n junction ($PN$) and p-i-n junction ($PIN$) based solar cells are numerically studied using the most realistic models, and optimized through mathematically rigorous multivariate optimization approaches. This analysis evidences optimal efficiencies of $17.8\%$ and $19.0\%$ for the $PN$ and $PIN$ structures. It also leads to propose, analyze and optimize p-layer free InGaN Schottky-Based Solar Cells (SBSC): the Schottky structure and a new $MIN$ structure for which the optimal efficiencies are shown to be a little higher than for the conventional structures: respectively $18.2\%$ and $19.8\%$.

The tolerance that is allowed on each parameter for each of the proposed cells has been studied. The new $MIN$ structure is shown to exhibit the widest tolerances on the layers thicknesses and dopings. In addition to its being p-layer free, this is another advantage of the $MIN$ structure since it implies its better reliability. Therefore, these new InGaN $SBSC$ are shown to be alternatives to the conventional structures that allow removing the p-type doping of InGaN while giving photovoltaic (PV) performances at least comparable to the standard multilayers $PN$ or $PIN$ structures.

\end{abstract}

\begin{keyword}

Simulation \sep Solar cell \sep InGaN \sep Schottky \sep SBSC


\PACS 61.72.uj \sep 85.60.-q \sep 88.40.hj


\end{keyword}

\end{frontmatter}

% -----------------------------------------------------------------------------------------

\section{Introduction}
\label{introduction}

The Indium Gallium Nitride (InGaN) ternary alloy has attracted attention as a potentially ideal candidate for high efficiency solar cells. Indeed, its bandgap can cover the whole solar spectrum, solely by changing its Indium composition \cite{bhuiyan2012ingan,reichertz_demonstration_2009}. The InGaN alloy also counts among its advantages a high absorption coefficient \cite{matioli2011high, lin2012simulation} as well as a good radiation tolerance \cite{polyakov2013radiation}, allowing its operation in extreme conditions.

However, one of its main drawbacks is the difficulty of its p-doping, owing mainly to the high residual donors' concentration, the lack of \emph{ad. hoc.} acceptors \cite{dahal_ingan_gan_2009} and the complex technological processes that are difficult to master and that highly impact the layer quality \cite{meng2010mg, gherasoiu2014ingan}.
The other drawbacks concern the difficulty to realize ohmic contacts \cite{bhuiyan2012ingan}, the poor InGaN material quality and the difficulty to grow InGaN with Indium content high enough to allow the optimal covering of the whole solar spectrum \cite{yamamoto_metal-organic_2013, durukan2015examination}.
For these reasons the InGaN based solar cell is still in early development stages and the reported PV efficiency is still very low to be competitive with other well established thin films technologies \cite{toledo_ingan_2012}.

That is the reason why we present a comprehensive comparative study of $PN$, $PIN$ and p-layer free Schottky Based Solar Cells ($SBSC$) structures using realistic physical models and rigorous mathematical optimization approaches and propose a new efficient p-layer free solar cell design with  performances higher and  tolerances wider than the previously studied Schottky structure \cite{ould_saad_hamady2016numerical}.

The following section \ref{modeling} describes the physical modeling and simulation methodology for the InGaN solar cell structures and discusses their main physical models and material parameters. Section \ref{results} presents the optimal results for the $PN$ and $PIN$ structures and discusses the impact of the p-layer parameters. Section \ref{SBSC} propose the replacement of the p-layer by a Schottky contact and discusses the performances of the resulting Schottky based solar cells, evidencing, in particular,  better fabrication tolerances for the new $MIN$ structure. Section \ref{actualInGaN} presents the results obtained using actual recently published InGaN experimental composition\cite{fabien2015large}, before section \ref{conclusion} concludes.

% -----------------------------------------------------------------------------------------

\section{Modeling and Simulation}
\label{modeling}

\subsection{Physical Modeling}
\label{physicalmodeling}

The physical modeling used throughout this paper to carry out the device simulations and optimizations presented in the next sections has been conceived with the less possible approximations and based, whenever possible, on actual measurements. It is summarized in this section.

% -----------------------------------------------------------------------------------------

\subsubsection{Transport modeling}
The mobilities for electrons and holes, needed for the drift-diffusion model, were calculated using the Caughey-Thomas expressions \cite{schwierz_electron_2005}:
\newcommand{\TRatio}{\left(\frac{T}{300}\right)}
\begin{equation}
\label{eq_mobility}
  \mu_m=\mu_{1m}\TRatio^{\alpha_m}+\frac{\mu_{2m}\TRatio^{\beta_m}- \mu_{1m}\TRatio^{\alpha_m}}{1+\left(\frac{N}{N_m^{\mathrm{crit}}\TRatio^{\gamma_m}}\right)^{\delta_m}}
,
\end{equation}
 where $m$ is either $n$ or $p$, $\mu_n$ being the electrons mobility and $\mu_p$ that of holes. $T$ is the absolute temperature. $N$ is the doping concentration. $N^{\mathrm{crit}}$ and the $n$ or $p$ subscripted $\alpha$, $\beta$, $\delta$ and $\gamma$ are the model parameters which depend on the Indium composition \cite{brown_finite_2010}.

In addition to the mobility model, were taken into account the bandgap narrowing effect \cite{schenk2008band}, as well as the Shockley–Read–Hall (SRH) \cite{ryu2009rate} and direct and Auger recombination models using the Fermi statistics \cite{bertazzi2010numerical}.

% -----------------------------------------------------------------------------------------

\subsubsection{Light absorption modeling}

Modeling InGaN based solar cells also implies  the need for a precise model of light absorption in the whole solar spectrum and for all $x$ Indium composition. We used a phenomenological model for InGaN that was proposed previously \cite{brown_finite_2010} as

\begin{equation}
  \Unit{\alpha}{cm^{-1}}=\Unit{10^5}{cm^{-1}}\sqrt{C\left(E_{ph}-E_g\right)+D\left(E_{ph}-E_g\right)^2},
  \label{eq_absorption}
\end{equation}
where $E_{ph}$ is the incoming photon energy, $E_g$ is the material bandgap at a given Indium composition, $C$ and $D$ are empirical parameters depending on the Indium composition.

For the refraction index we used the Adachi model \cite{djurisic_modeling_1999}, defined for InGaN and for a given photon energy as
\newcommand{\ERatio}{\frac{E_{ph}}{E_g}}
\begin{equation}
  n\left(E_{ph}\right)=\sqrt{\frac{A}{\left(\ERatio\right)^2}\left[2-\sqrt{1+\ERatio}-\sqrt{1-\ERatio}\right]+B },
  \label{eq_index}
\end{equation}
where $A$ and $B$ are also empirical parameters depending on the Indium composition.

\subsubsection{Material parameters}

\begin{table}

\begin{center}
\subtable[Data  from ref \cite{brown_finite_2010}.]{
\begin{tabular}{|c|c|c|c|c|c|}
\hline
~ & $\Unit{E_g}{eV}$ & $\Unit{\chi}{eV}$ & $\Unit{N_c}{cm^{-3}}$ & $\Unit{N_v}{cm^{-3}}$ & $\varepsilon$ \\ \hline
GaN & 3.42 & 4.1 & $\E{2.3}{18}$ & $\E{4.6}{19}$ & 8.9 \\ \hline
InN & 0.7 & 5.6 & $\E{9.1}{17}$ & $\E{5.3}{19}$ & 15.3  \\ \hline
    \end{tabular}

\label{tab_physparam}
} 

\subtable[Data from refs \cite{nawaz_tcad_2012,wang2014analytical}.]
{\begin{tabular}{|c|c|c|c|c|c|c|c|}
\hline
~ &  $\Unit{\mu_n^1}{cm^2/Vs}$ &  $\Unit{\mu_n^2}{cm^{2}/Vs}$   & $\delta_n$ & $\Unit{N_n^{\mathrm{crit}}}{cm^{-3}}$ \\ \hline
GaN & 295  & 1460  & 0.71 & $\E{7.7}{16}$\\ \hline
InN & 1982.9  & 10885  & 0.7439 & $\E{1.0}{17}$ \\ \hline
    \end{tabular}
\label{tab_mobilparam_n}}

\subtable[Data from ref \cite{brown_finite_2010}.]
{\begin{tabular}{|c|c|c|c|c|c|c|c|}
\hline
~ &  $\Unit{\mu_p^1}{cm^2/Vs}$ &  $\Unit{\mu_p^2}{cm^{2}/Vs}$   & $\delta_p$ & $\Unit{N_p^{\mathrm{crit}}}{cm^{-3}}$ \\ \hline
GaN & 3.0  & 170  & 2.0 & $\E{1.0}{18}$\\ \hline
InN & 3.0  & 340  & 2.0 & $\E{8.0}{17}$ \\ \hline
    \end{tabular}
\label{tab_mobilparam_p}}
\end{center}
\caption{Experimental or \emph{ab initio} data used in the simulations. Owing to the absence of any experimental data, $\alpha_n$, $\beta_n$, $\gamma_n$, $\alpha_p$, $\beta_p$ and $\gamma_p$ have been estimated to 1. \label{tab_param}}
\end{table}

The material dependent parameters have been determined for GaN and InN binaries, either from experimental work or \emph{ab initio} calculations \cite{nawaz_tcad_2012,brown_finite_2010}. A review of their values is given in Table \ref{tab_param}.

In the following, the values for the material parameters of InGaN, for any Indium composition $x\in\left[0,1\right]$, were linearly interpolated in between the GaN and InN binaries, except for the bandgap $E_g$ and the electronic affinity $\chi$ where we used the modified Vegard Law with a bowing factor $b=1.43 \mathrm{eV}$ for the bandgap and $b=0.8 \mathrm{eV}$ for the affinity \cite{franssen2008bowing,brown_finite_2010} respectively.

\begin{table}
  \begin{center}
\begin{tabular}{|l|c|c|}
\hline
Indium Composition & $\Unit{C}{eV^{-1}}$ & $\Unit{D}{eV^{-2}}$ \\ \hline
1	&0.69642	&0.46055 \\ \hline
0.83	&0.66796	&0.68886 \\ \hline
0.69	&0.58108	&0.66902 \\ \hline
0.57	&0.60946	&0.62182 \\ \hline
0.5	&0.51672	&0.46836 \\ \hline
0	&3.52517	&-0.65710 \\ \hline
  \end{tabular}
  \end{center}
\caption{Values for $C$ and $D$ in equation (\ref{eq_absorption}) as found by Brown \emph{et. al.} in \cite{brown_finite_2010}.}
\label{tab_Brown}
\end{table}

For the recombination models, we chose a relatively low carrier lifetime value of 1ns, much lower than the value of 40ns reported for GaN \cite{vscajev2013time}, in order to get as realistic results as can be.

For the light absorption model, the values of $C$ and $D$ in equation (\ref{eq_absorption}) are taken from the experimental measurement reported in \cite{brown_finite_2010} and summarized in Table \ref{tab_Brown}. We approximated their dependency on the Indium composition $\X$ by a polynomial fit, of the $\mathrm{4}^{th}$ degree for the former, and quadratic for the latter:

\begin{eqnarray*}
C &=& 3.525 - 18.29\X + 40.22\X^2 - 37.52\X^3 + 12.77\X^4,\\
D &=& -0.6651 + 3.616\X - 2.460\X^2.
\end{eqnarray*}

The $A$ and $B$ parameters in the refraction index equation (\ref{eq_index}) are experimentally measured \cite{nawaz_tcad_2012,brown_finite_2010} for GaN ($A^{\mathrm{GaN}}=9.31$ and $B^{\mathrm{GaN}}=3.03$) and InN ($A^{\mathrm{InN}}=13.55$ and $B^{\mathrm{InN}}=2.05$) and linearly interpolated for InGaN.

Finally, we have chosen to shine on the cell the ASTM-G75-03 solar spectrum taken from the National Renewable Energy Laboratory database\footnote{\url{http://rredc.nrel.gov/solar/spectra/am1.5/astmg173/astmg173.html}}.

% -----------------------------------------------------------------------------------------

\subsection{Simulation Methodology}
\label{simulation}

The devices are simulated in the framework of a drift-diffusion model using the \Atlas device simulation software from the \Silvaco  suite, in which we implemented our physical models.
\Atlas solves, in two dimensions, the drift-diffusion nonlinear partial differential problem using the Newton coupled and the Gummel decoupled methods \cite{selberherr2012analysis}. The solar cell analyzed characteristics were the spectral response, the I-V characteristics, the inner electric field and potential distributions as well as the recombination rate variations.

We used mathematically rigorous multivariate optimization methods to find the optimum efficiency with respect to a given set of parameters (as later shown in tables \ref{tab1_Optimum} and \ref{tab2_Optimum}). This methodology is far more rigorous than the usual single parametric analysis, where one parameter is varying while the other parameters are kept constant. It yields for instance the absolute optimum efficiency as a function of the physical parameters. We have used three mathematical optimization methods that give very similar results within a comparable amount of computing time (typically a few hours per simulation with a highly optimized code): the truncated Newton algorithm (TNC) \cite{nocedal2006large}, the Sequential Least SQuares Programming (SLSQP) method \cite{kraft_software_1988} and the L-BFGS-B quasi-Newton method \cite{nocedal2006large}.
The optimization work has been done with a Python package we developed in the SAGE \cite{sage} interface to the SciPy \cite{van_der_walt_numpy_2011,scipy} optimizers, using the \Atlas simulator as the backend engine.

% -----------------------------------------------------------------------------------------

\section{Optimization of PN and PIN structures}
\label{results}

\subsection{Optimal Results}

\begin{figure}
  \subfigure[PN InGaN based solar cell.]{
	\includegraphics[width=\linewidth]{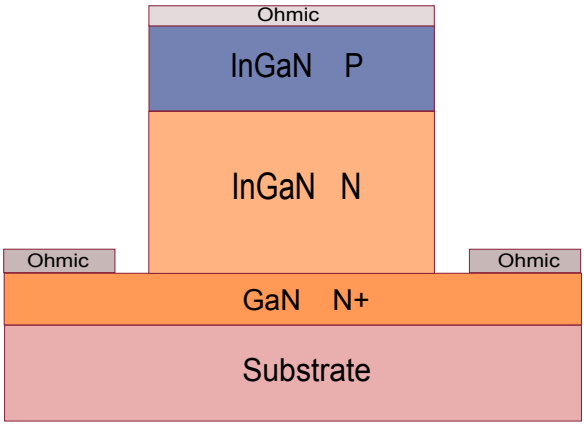}
	\label{PN}
 }
\subfigure[PIN InGaN based solar cell, where the the quasi-intrinsic layer is slightly n-doped.]{
	\includegraphics[width=\linewidth]{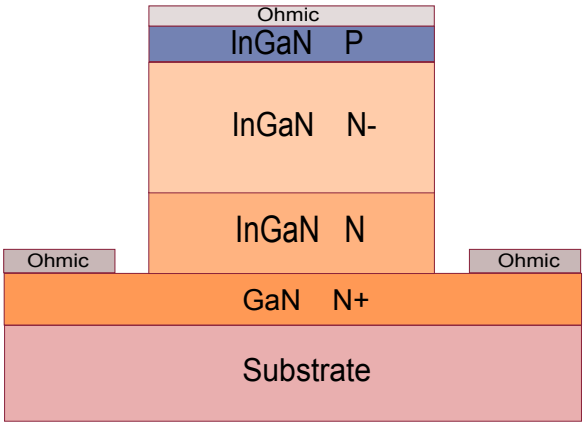}
	\label{PIN}
}
\caption{Schematic views of the InGaN based PN and PIN solar cells structures.}
\label{PN_PIN}
\end{figure}
	
\begin{table*}
	\scriptsize
	\centering
	\setlength{\tabcolsep}{4pt}
	\renewcommand{\arraystretch}{1.0}
	\begin{tabular}{|>{\bfseries}l|c|c|c|c|c|c|c|c|}
	
	\cline{2-9}\multicolumn{1}{c|}{} &  &  &  &  &  &  &  & $\bm{\eta(\%)}$ \\

	\multicolumn{1}{c|}{} & $\bm{\Lp(\mu m)}$ & $\bm{\Li(\mu m)}$ & $\bm{\Ln(\mu m)}$ & $\bm{\Na(cm^{-3})}$ & $\bm{\Ni(cm^{-3})}$ & $\bm{\Nd(cm^{-3})}$ & $\bm{x}$ &  $\bm{V_{OC}(V)}$ \\
	
	\multicolumn{1}{c|}{} &  &  &  &  &  &  &  & $\bm{J_{SC}(mA/cm^2)}$ \\

	\multicolumn{1}{c|}{} &  &  &  &  &  &  &  & $\bm{FF(\%)}$ \\

	\hline
	
	Range & $\bm{[0.10 - 1.00]}$ & $\bm{[0.10 - 1.00]}$ & $\bm{[0.10 - 1.00]}$ & $\bm{[1.0\times10^{16} - 1.0\times10^{19}]}$ & $\bm{[1.0\times10^{14} - 1.0\times10^{17}]}$ & $\bm{[1.0\times10^{16} - 1.0\times10^{19}]}$ & \shortstack[c]{\\ $\bm{[0.00 - 1.00]}$} & \\
	
	\hline
	
	   &  & \cellcolor[gray]{.9} &  &  & \cellcolor[gray]{.9} &  &  & $\bm{17.8}$ \\

	PN & $\bm{0.01}$ & \cellcolor[gray]{.9} & $\bm{1.00}$ & $\bm{1.0\times10^{19}}$ & \cellcolor[gray]{.9} & $\bm{3.9\times10^{16}}$ & $\bm{0.56}$ & $\bm{0.855}$ \\

	   & $\bm{[0.01 - 0.04]}$ & \cellcolor[gray]{.9} & $\bm{[0.48 - 1.00]}$ & $\bm{[4.4\times10^{16}-1.0\times10^{19}]}$ & \cellcolor[gray]{.9} & $\bm{[1.0\times10^{16}-3.5\times10^{17}]}$ & $\bm{[0.50-0.72]}$ & $\bm{26.75}$ \\

	   &  & \cellcolor[gray]{.9} &  &  & \cellcolor[gray]{.9} &  &  & $\bm{77.85}$ \\

	\hline
	
	   &  &  &  &  &  &  &  & $\bm{19.0}$ \\

 PIN  & $\bm{0.01}$ & $\bm{0.54}$ & $\bm{0.50}$ & $\bm{1.0\times10^{19}}$ & $\bm{5.8\times10^{16}}$ & $\bm{5.0\times10^{17}}$ & $\bm{0.59}$ & $\bm{0.875}$ \\
	
      & $\bm{[0.01 - 0.04]}$ & $\bm{[0.18 - 1.00]}$ & $\bm{[0.10 - 1.00]}$ & $\bm{[5.9\times10^{16}-1.0\times10^{19}]}$ & $\bm{[1.0\times10^{14}-1.0\times10^{17}]}$ & $\bm{[1.9\times10^{16}-1.0\times10^{19}]}$ & $\bm{[0.47-0.71]}$ & $\bm{27.36}$ \\

	  &  &  &  &  &  &  &  & $\bm{79.39}$ \\

	\hline
	
	\end{tabular}
	
	\caption{Optimum efficiency $\eta$ obtained for the $PN$ and $PIN$ structures and associated open-circuit voltage $V_{OC}$, short-circuit current $J_{SC}$ and Fill Factor $FF$, along with the corresponding physical and material parameters. These results are obtained from several optimizations with random starting points ensuring the absoluteness of the optimum efficiency $\eta$. $\X$ is the Indium composition. $\Lp$, $\Li$ and $\Ln$ are the thicknesses of the $P$, $I$ and $N$ layers respectively and where applicable. $\Na$, $\Ni$ and $\Nd$ are the dopings of the $P$, $I$ and $N$ layers respectively where applicable. For each parameter, a range and a tolerance range are given. The range is on the second line of the table. It is the range within which the optimum value of a given parameter is sought. The tolerance range is given just below each parameter optimal value. It corresponds to the set of values of that parameter for which the efficiency $\eta$ remains above $90\%$ of its maximum, the other parameters being kept at their optimum values. }
   \label{tab1_Optimum}
\end{table*}

The $PN$ and $PIN$ solar cells are schematically shown in Figure \ref{PN_PIN}. These devices could be realized in practice using the developed growth techniques of InGaN on GaN/sapphire substrates and the device realization techniques \cite{yamaguchi2013growth}.

The first device, shown on figure \ref{PN}, is an InGaN $PN$ structure. Its optimization, and eventually its practical realization with a competitive efficiency, is the \emph{sine qua none} condition to actually manufacturing the high efficiency multijunction next-generation solar cells \cite{zhang2007simulation, fischer_highly_2013, fabien_guidelines_2014}. The physical parameters for which the optimum was sought are shown on table \ref{tab1_Optimum}: the relevant five parameters for the $PN$ structure are the thickness and dopings of the two layers, along with their common Indium composition.

The second design is based on a $PIN$ structure where the "intrinsic" layer consists in an n-doped layer with a relatively low doping concentration. The standard intrinsic layer (i-layer) has been replaced by a slightly doped n-layer for two reasons: on the one hand, the elaborated InGaN usually exhibits residual n-doping \cite{pantha2011origin, hoffbauer2013rich} and, on the other hand as we will demonstrate later in this section, the optimal efficiency for a $PIN$ solar cell is obtained for an intermediate n-doped layer and not for the quasi-intrinsic layer. The resulting structure is shown on figure \ref{PIN}. The seven optimization parameters are shown in table \ref{tab1_Optimum}, with the thickness and dopings of the three layers along with their common Indium composition. The minimum value of the quasi-intrinsic layer doping has been set lower than the usually reported residual doping value in InGaN \cite{pantha2011origin}.

To optimize these devices, we used the mathematical optimization methods presented in section \ref{simulation}. These methods are constrained and therefore need a parameter range, and, as for the non-constrained methods, a starting point. We defined the parameter range to ensure the physical meaning and the technological feasibility of each parameter. The range chosen for each parameter is shown on the second line of table \ref{tab1_Optimum}. We then chose to run the optimization with several randomly chosen starting points to get an insight into the precision of our computation and ensure that the found optimum is absolute.

For the $PN$ structure, we found a maximum cell efficiency of $17.8\%$ and optimal values for the physical parameters. The optimal thickness of
 the $P$ layer is found to be $\Lp=0.01 \mathrm{\mu m}$. That of  the $N$ layer  thickness  is $\Ln=1.00 \mathrm{\mu m}$. The optimal doping of the $P$ layer is $\Na=1.0\times10^{19} \mathrm{cm^{-3}}$. That of the $N$ layer is $\Nd=3.9\times10^{16} \mathrm{cm^{-3}}$. The optimal Indium composition is $\X=0.56$. The corresponding open-circuit voltage ($V_{OC}$) is $0.855 \mathrm{V}$ with a short-circuit current ($J_{SC}$) of $26.75 \mathrm{mA/cm^{2}}$ and a fill factor ($FF$) of $77.85\%$.

For the $PIN$ structure, we found a maximum cell efficiency of $19.0\%$ for the optimal values of the following parameters:
a $P$ layer thickness of $\Lp=0.01 \mathrm{\mu m}$, an  $I$ layer thickness of  $\Li=0.54 \mathrm{\mu m}$, a $N$ layer thickness of  $\Ln=0.50 \mathrm{\mu m}$, a $P$ layer doping of  $\Na=1.0\times10^{19} \mathrm{cm^{-3}}$, an $I$ layer doping of  $\Ni=5.8\times10^{16} \mathrm{cm^{-3}}$, a $N$ layer doping of  $\Nd=5.0\times10^{17} \mathrm{cm^{-3}}$ and an Indium composition of $\X=0.59$. The corresponding PV parameters are $V_{OC}=0.875 \mathrm{V}$, $J_{SC}=\mathrm{27.36 mA/cm^{2}}$ and $FF=79.39\%$.

All these parameters with their tolerance range, as defined below, are reported in table \ref{tab1_Optimum}.

In practice, it is indeed necessary for an optimal parameter to have a wide \emph{tolerance range} in which it can vary without lowering  the cell efficiency too much. We have performed the \emph{tolerance} analysis on each parameter, while keeping all the others at their optimal value. We have thus defined a tolerance range, which is the range of values of a given parameter for which the efficiency $\eta$ remains above $90\%$ of its maximum value. The tolerance range is shown on table \ref{tab1_Optimum}, just below the optimal value. For instance, for the $PN$ structure, the efficiency value remains between $16.0\%$ and $17.8\%$ for a p-layer doping $\Na$ varying between $4.4\times10^{16} \mathrm{cm^{-3}}$ and $1.0\times10^{19} \mathrm{cm^{-3}}$, the other parameters remaining at their optimal values. Table \ref{tab1_Optimum} shows that, on the one hand, the $PIN$ solar cell has an efficiency slightly higher than that of the $PN$ solar cell and, on the other hand, the tolerance ranges for layers thicknesses in the $PIN$ structure are wider than in the $PN$ structure. This latter property is a considerable advantage of the $PIN$ structure in the practical cell realization. For instance, the $PIN$ structure has a tolerance range of $[0.10 - 1.00] \mathrm{\mu m}$ for the n-layer thickness, almost twice wider than that of the $PN$ structure. The wider tolerance range for the n-doping in the $PIN$ structure also allows increasing the n-layer doping value without noticeably impacting the efficiency, for designing low resistance ohmic contacts \cite{jani2007design}.

% -----------------------------------------------------------------------------------------

\subsection{Impact of the p-layer parameters}
\label{impactPLayer}

\begin{figure}
  
\subfigure[InGaN PN solar cell efficiency as a function of the p-layer doping with various thicknesses. The other parameters are kept at their optimal value.]{
	\includegraphics[width=\linewidth]{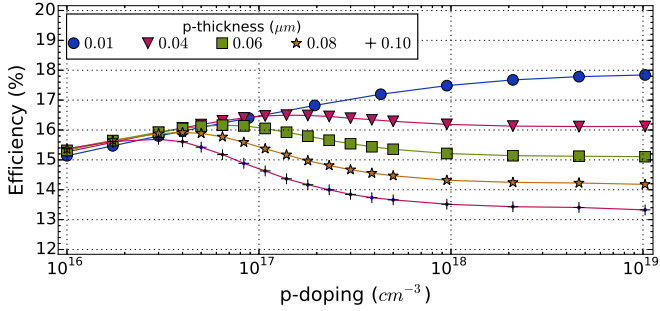}
	\label{PN_player_eta}
}
\subfigure[InGaN PN solar cell optimal p-layer doping variation with thickness. The efficiency corresponding to some points are shown.]{
	\includegraphics[width=\linewidth]{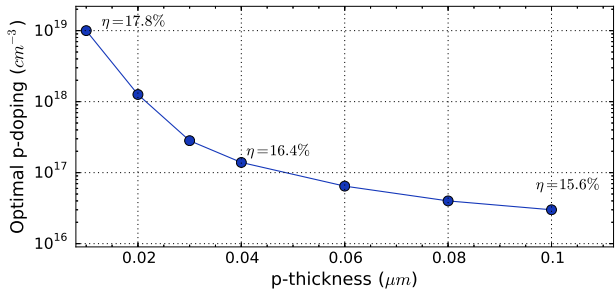}
	\label{PN_player_dop}
}

\subfigure[InGaN PN solar cell current-voltage characteristics for different p-layer doping/thickness optimal couples corresponding to points shown in figure \ref{PN_player_dop}.]{
	\includegraphics[width=\linewidth]{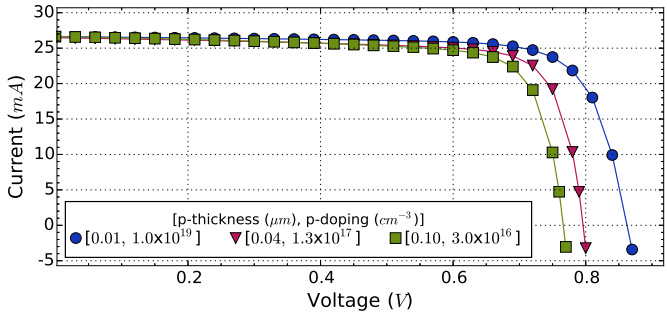}
	\label{PN_player_jv}
}
 
\caption{Impact of the p-layer parameters in the optimal InGaN PN solar cell electrical characteristics. The optimal parameters are given in table \ref{tab1_Optimum}. All 3 sub-figures are drawn from a set of calculated points approximately 10 times larger than actually displayed. This allow better readability. For each curve, the hidden points lie on the displayed line.}
\label{Curves_PN}
\end{figure}

\begin{figure}
\subfigure[InGaN PIN solar cell efficiency as a function of the p-layer doping with various thicknesses. The other parameters are kept at their optimal value.]{
	\includegraphics[width=\linewidth]{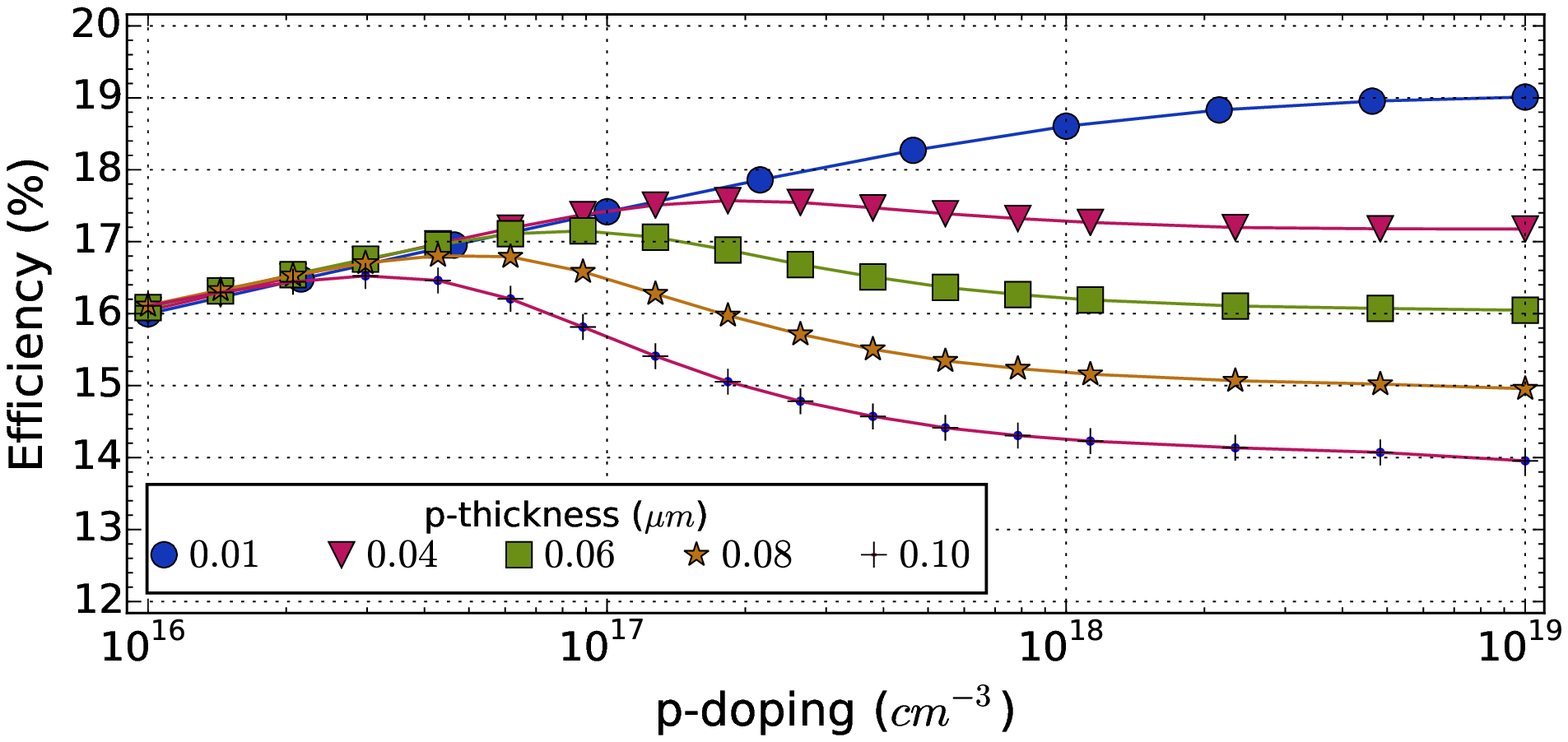}
	\label{PIN_player_eta}
}
\subfigure[InGaN PN optimal p-doping variation with thickness. The efficiency corresponding to some points are shown.]{
	\includegraphics[width=\linewidth]{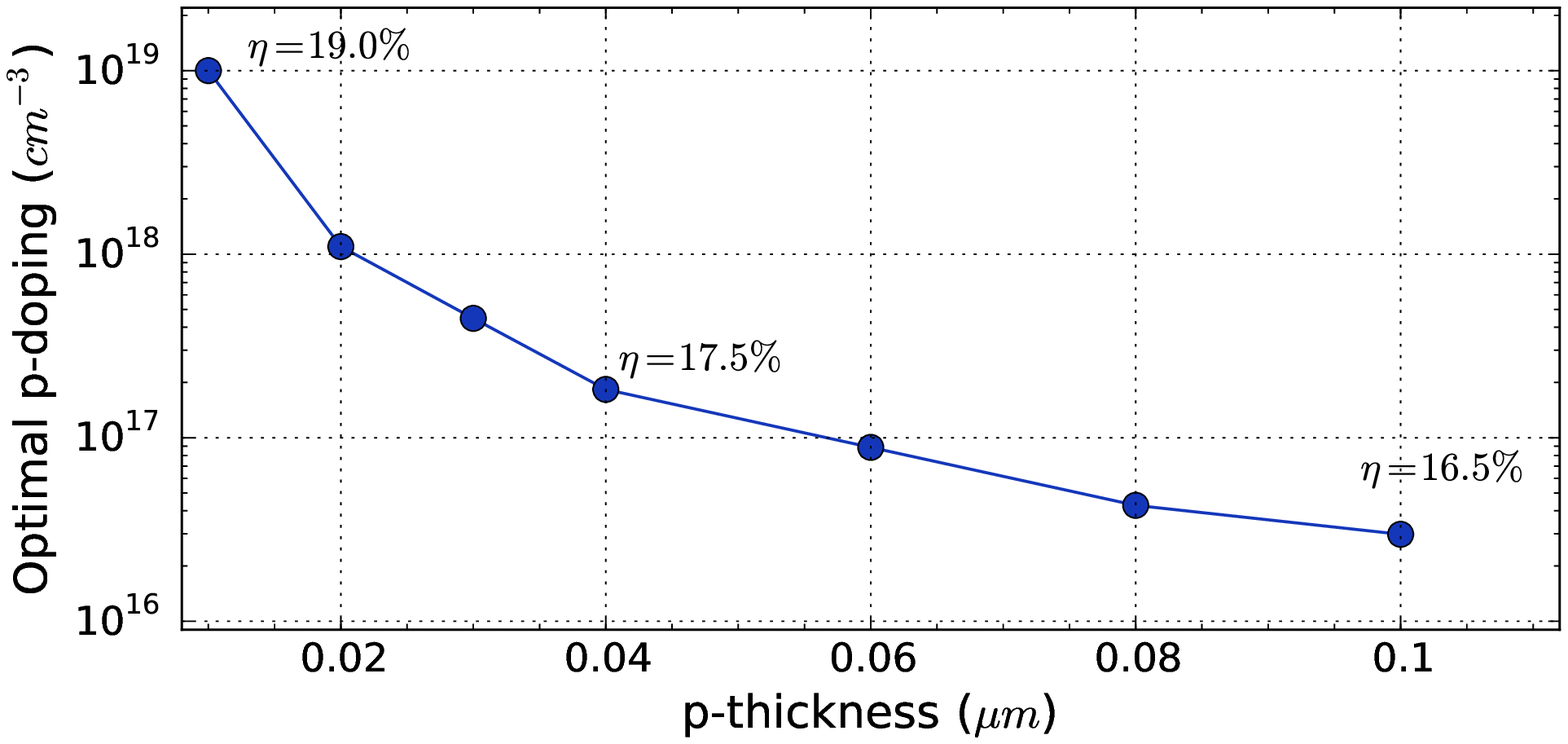}
	\label{PIN_player_dop}
}

\subfigure[InGaN PIN solar cell current-voltage characteristics or different p-layer doping/thickness optimal couples corresponding to points shown in figure \ref{PIN_player_dop}.]{
	\includegraphics[width=\linewidth]{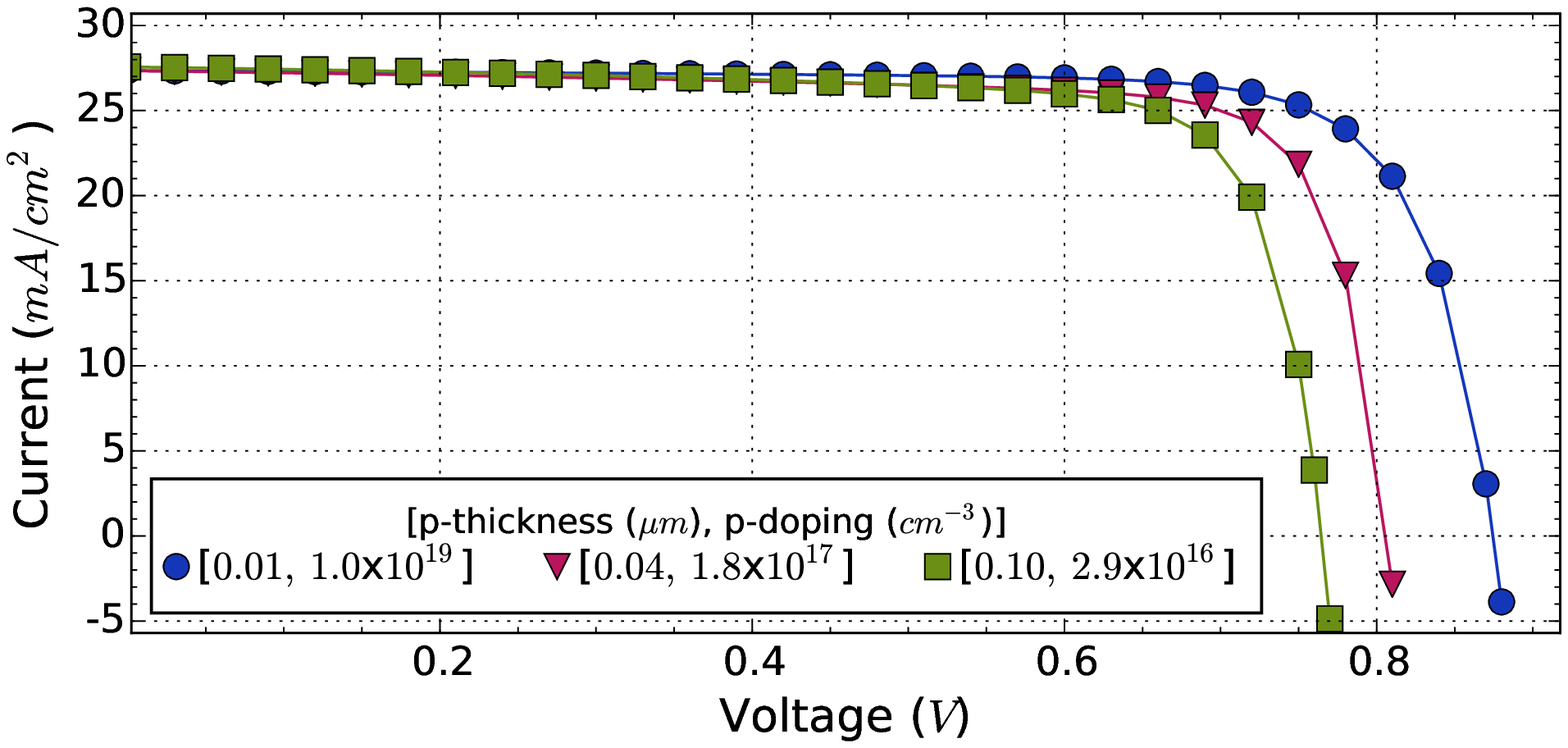}
	\label{PIN_player_jv}
}
 
\caption{Impact of the p-layer parameters in the optimal InGaN PIN solar cell electrical characteristics. The optimal parameters are given in table \ref{tab1_Optimum}. All 3 sub-figures are drawn from a set of calculated points approximately 10 times larger than actually displayed. This allow better readability. For each curve, the hidden points lie on the displayed line.}
\label{Curves_PIN}
\end{figure}

Figures \ref{PN_player_eta} and \ref{PIN_player_eta} show the efficiency as a function of the p-layer doping for various p-layer thicknesses (p-thicknesses) for $PN$ and $PIN$ structures respectively. These results show that the efficiency is optimal for a given p-doping which increases when the p-thickness decreases.

Figures \ref{PN_player_dop} and \ref{PIN_player_dop} display the optimal p-doping variation with the p-thicknesses for the $PN$ and $PIN$ structures respectively. For the $PN$ structure, the corresponding efficiency varies from $17.8\%$ down to $15.6\%$ with the thickness of the p-layer varying from $0.01\mu m $ to $0.10\mu m $. The corresponding efficiency for the $PIN$ structure, varies from $19.8\%$ down to $16.5\%$ with the thickness of the p-layer varying from $0.01\mu m $ to $0.10\mu m $.

Figures \ref{PN_player_jv} and \ref{PIN_player_jv} display the I-V characteristics for some p-thicknesses values, for the $PN$ and $PIN$ structures respectively. The I-V curves for the $PN$ structure were obtained for the thicknesses of the p-layer of $0.01\micron$, $0.04\micron$ and $0.10\micron$ corresponding to the optimal p-doping values of $1.0\times10^{19}\mathrm{cm^{-3}}$, $1.3\times10^{17}\mathrm{cm^{-3}}$ and $3.0\times10^{16}\mathrm{cm^{-3}}$ respectively. 
For the $PIN$ structure, the I-V curves were obtained for the same thicknesses corresponding to the optimal p-doping values of $1.0\times10^{19}\mathrm{cm^{-3}}$, $1.8\times10^{17}\mathrm{cm^{-3}}$ and $2.9\times10^{16}\mathrm{cm^{-3}}$ respectively.
For both the $PN$ and the $PIN$ structures, the short-circuit current $J_{SC}$ remains almost constant while the open-circuit voltage $V_{OC}$ increases when decreasing the p-thickness and increasing the p-doping along the optimal curve of figures \ref{PN_player_dop} and \ref{PIN_player_dop}. 

For a given thickness, say $0.10 \mu m$, the maximum electric field value obviously increases with the p-doping while the space charge region (SCR) width decreases, as well as the recombination rate. This is mainly due to the SRH recombination mechanism. These two variations lead to an increase of the $V_{OC}$ and, in the same time, to a decrease of $J_{SC}$. These two opposing trends of $J_{SC}$ and $V_{OC}$ lead to maximum efficiency points depending on the p-doping and thickness of the p-layer, as shown in Figures \ref{PN_player_eta} and \ref{PIN_player_eta}.

Figures \ref{PN_player_dop} and \ref{PIN_player_dop} summarize this variation showing that even if the absolute optimal doping is high, the efficiency remains relatively high for wide doping and thickness ranges. For instance, for the $PIN$ structure, a p-thickness of $0.04 \mu m$ and a p-doping of $1.8\times10^{17} \mathrm{cm^{-3}}$ lead to an efficiency of $17.5\%$. It remains relatively close to the optimal one obtained for a thickness of $0.01 \mu m$ and a doping of $1.0\times10^{19} \mathrm{cm^{-3}}$. This point concerning the tolerance range, as previously underlined for $PN$ and $PIN$ solar cells, is of great importance for the practical solar cell realization and it will be discussed for $SBSC$ in the following section.

Figures \ref{PN_player_jv} and \ref{PIN_player_jv} show that the efficiency variation with the p-layer thickness and doping is mainly due to variations in $V_{OC}$. Indeed, the short-circuit current $J_{SC}$ remains almost constant owing to increasing p-doping associated to decreasing thickness, whereas $V_{OC}$ increases owing to increasing p-doping only.

All these results show that the optimal performances of both the $PN$ and $PIN$ structures are obtained for a relatively thin p-layer ($10$ $nm$) with a relatively high p-doping value of $1.0\times10^{19}\mathrm{cm^{-3}}$. Considering on the one hand that the optimal thickness is much lower than the mean penetration depth and diffusion length in InGaN and on the other hand that the optimal doping is relatively high, we propose an alternative that allows the removal of the p-layer. These alternatives are Schottky Based Solar Cells (SBSC), which correspond on the one hand to a Schottky junction and, on the other hand to a new structure.

% -----------------------------------------------------------------------------------------

\section{Schottky Based Solar Cells}
\label{SBSC}

\begin{figure}
\subfigure[InGaN based Schottky solar cell.]{
	\includegraphics[width=\linewidth]{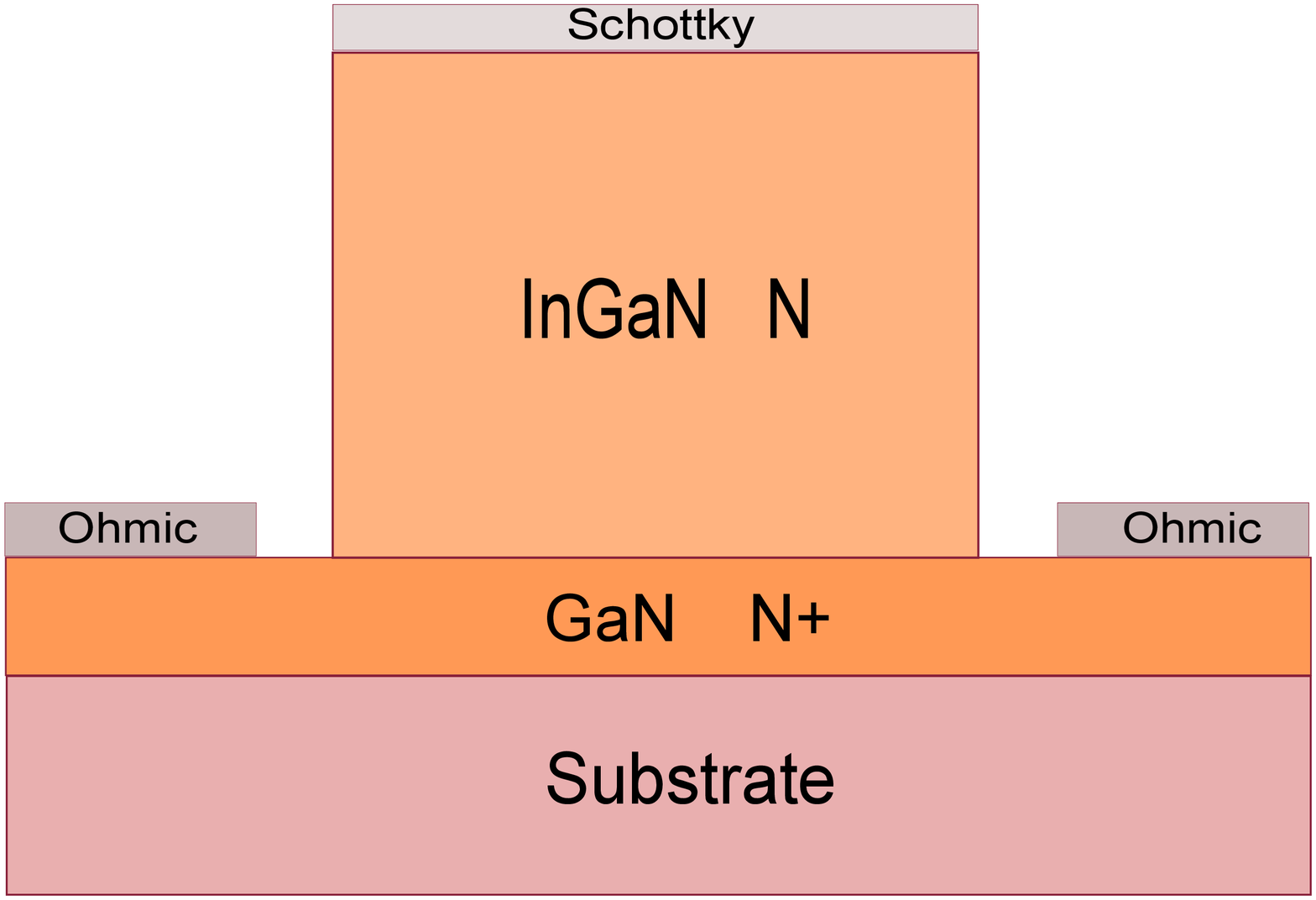}
	\label{SCH}
}
\subfigure[InGaN based MIN solar cell.]{
	\includegraphics[width=\linewidth]{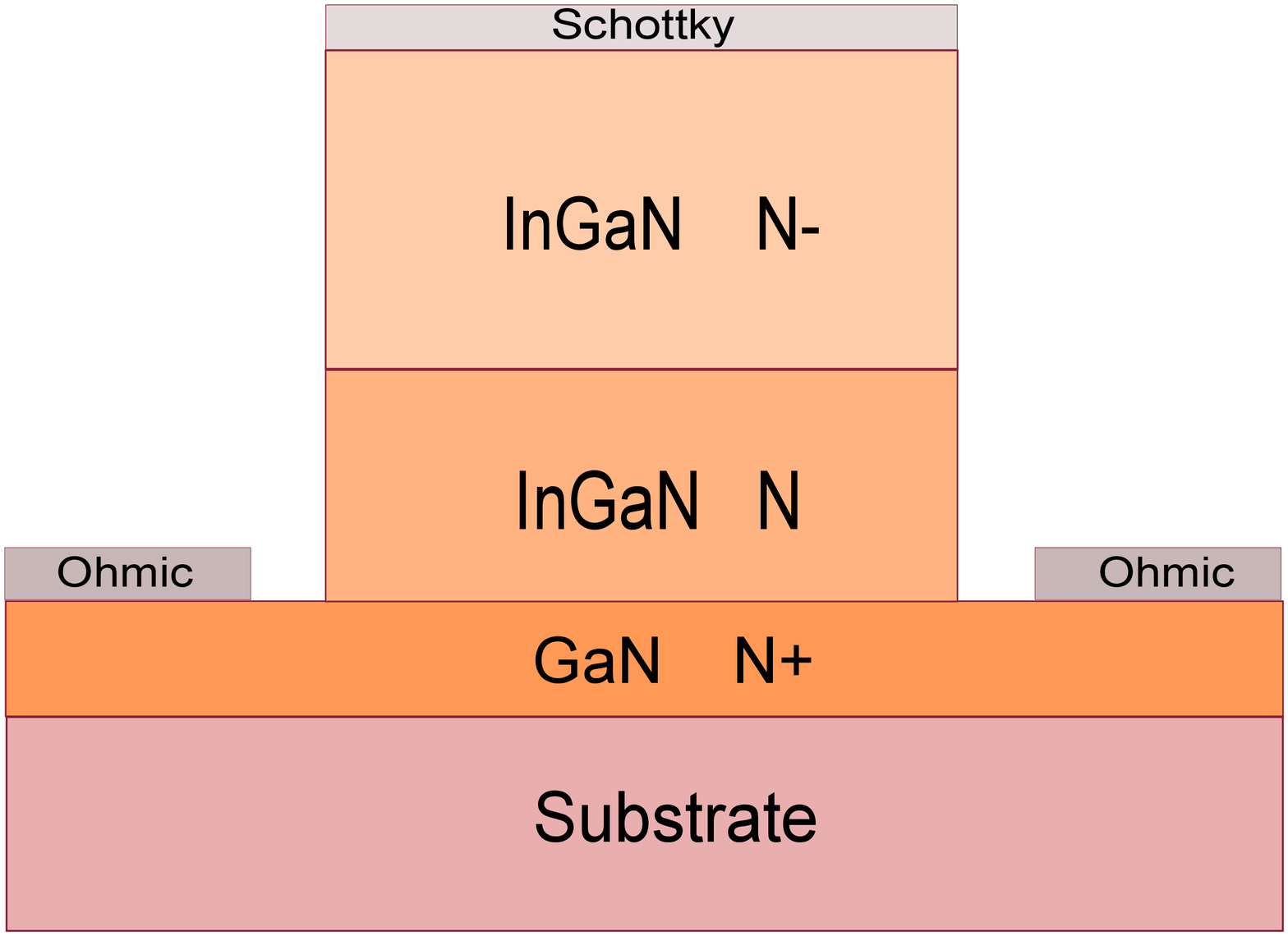}
	\label{MIN}
} 
\caption{Schematic views of the Schottky and MIN solar cells structures.}
\label{SCH_MIN}
\end{figure}

\begin{table*}
	\footnotesize
	\begin{center}

	\begin{tabular}{|>{\bfseries}l|c|c|c|c|c|c|c|}

	\cline{2-8}\multicolumn{1}{c|}{} &  &  &  &  &  &  & $\bm{\eta(\%)}$ \\

	\multicolumn{1}{c|}{} & $\bm{\Li(\mu m)}$ & $\bm{\Ln(\mu m)}$ & $\bm{\Ni(cm^{-3})}$ & $\bm{\Nd(cm^{-3})}$ & $\bm{\WF(eV)}$ & $\bm{x}$ &  $\bm{V_{OC}(V)}$ \\

	\multicolumn{1}{c|}{} &  &  &  &  &  &  & $\bm{J_{SC}(mA/cm^2)}$ \\

	\multicolumn{1}{c|}{} &  &  &  &  &  &  & $\bm{FF(\%)}$ \\

	\hline
		
		Range & $\bm{[0.10 - 1.00]}$ & $\bm{[0.10 - 1.00]}$ & $\bm{[1.0\times10^{14} - 1.0\times10^{17}]}$ & $\bm{[1.0\times10^{16} - 1.0\times10^{19}]}$ & $\bm{[5.50 - 6.30]}$ & \shortstack[c]{\\ $\bm{[0.00 - 1.00]}$} & \\

	\hline
	
	       & \cellcolor[gray]{.9} &  & \cellcolor[gray]{.9} &  &  &  & $\bm{18.2}$ \\
	
 Schottky & \cellcolor[gray]{.9} & $\bm{0.86}$ & \cellcolor[gray]{.9} & $\bm{6.5\times10^{16}}$ & $\bm{6.30}$ & $\bm{0.56}$ & $\bm{0.863}$ \\

	       & \cellcolor[gray]{.9} & $\bm{[0.53 - 1.00]}$ & \cellcolor[gray]{.9} & $\bm{[1.0\times10^{16}-3.0\times10^{17}]}$ & $\bm{[6.15-6.30]}$ & $\bm{[0.50-0.72]}$ & $\bm{26.80}$ \\

	       & \cellcolor[gray]{.9} &  & \cellcolor[gray]{.9} &  &  &  & $\bm{78.82}$ \\

	\hline

	       &  &  &  &  &  &  & $\bm{19.8}$ \\
	
      MIN & $\bm{0.61}$ & $\bm{0.83}$ & $\bm{6.1\times10^{16}}$ & $\bm{3.6\times10^{17}}$ & $\bm{6.30}$ & $\bm{0.60}$ & $\bm{0.835}$ \\

	       & $\bm{[0.10-1.00]}$ & $\bm{[0.10 - 1.00]}$ & $\bm{[1.0\times10^{14}-1.0\times10^{17}]}$ & $\bm{[1.8\times10^{16}-1.0\times10^{19}]}$ & $\bm{[6.11-6.30]}$ & $\bm{[0.48-0.72]}$ & $\bm{30.29}$ \\

	       &  &  &  &  &  &  & $\bm{78.39}$ \\

	\hline

	\end{tabular}
	\end{center}
	\caption{Optimum efficiency $\eta$ obtained for the $Schottky$ and $MIN$ structures and associated open-circuit voltage $V_{OC}$, short-circuit current $J_{SC}$ and Fill Factor $FF$, along with the corresponding physical and material parameters. These results are obtained from several optimizations with random starting points ensuring the absoluteness of the optimum efficiency $\eta$. $\X$ is the indium composition. $\Li$ and $\Ln$ are the thicknesses of the $I$ and $N$ layers respectively and where applicable. $\Ni$ and $\Nd$ are the dopings of the $I$ and $N$ layers respectively where applicable. For each parameter, a range and a tolerance range are given. The range is on the second line of the table. It is the range within which the optimum value of a given parameter is sought. The tolerance range is given just below each parameter optimal value. It corresponds to the set of values of that parameter for which the efficiency $\eta$ remains above $90\%$ of its maximum, the other parameters being kept at their optimum values. }
	\label{tab2_Optimum}
\end{table*}

\begin{figure}
	\includegraphics[width=\linewidth]{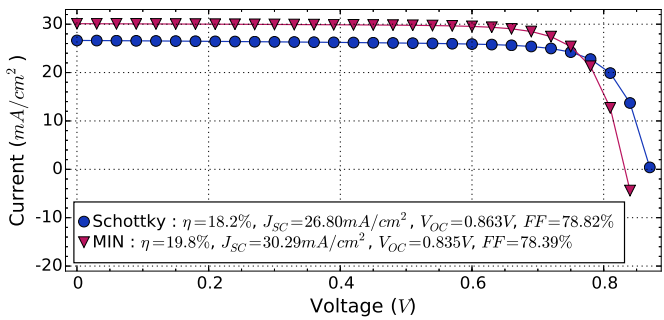}
	\caption{Current-voltage characteristics for the $InGaN$ Schottky and MIN solar cells.}
\label{sbsc_jv}
\end{figure}

\begin{figure}
  \subfigure[MIN PV efficiency as a function of the i-layer thickness.]{
	\includegraphics[width=\linewidth]{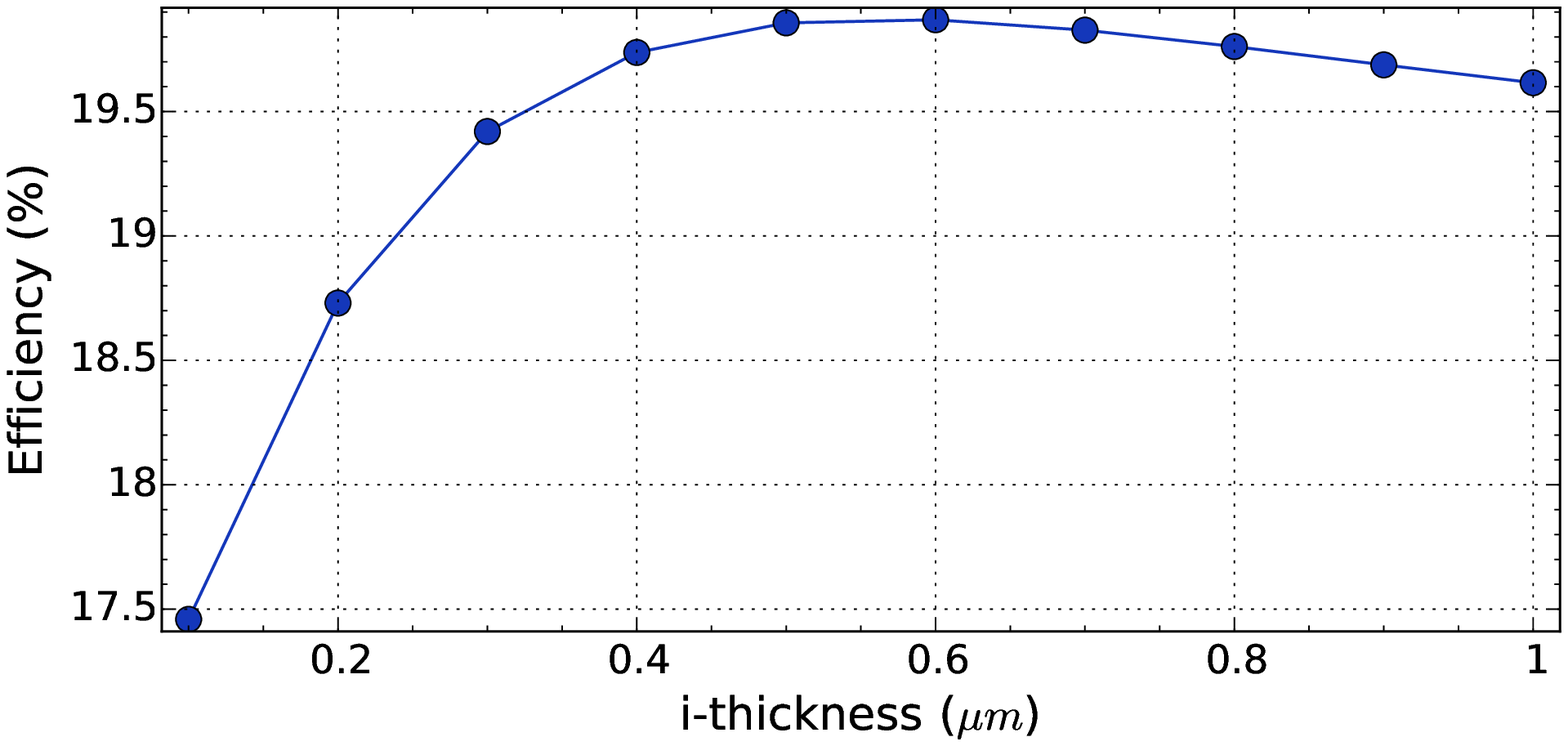}
	\label{min_eff_thick}
 }
\subfigure[MIN PV efficiency as a function of the i-doping for various i-layer thicknesses.]{
	\includegraphics[width=\linewidth]{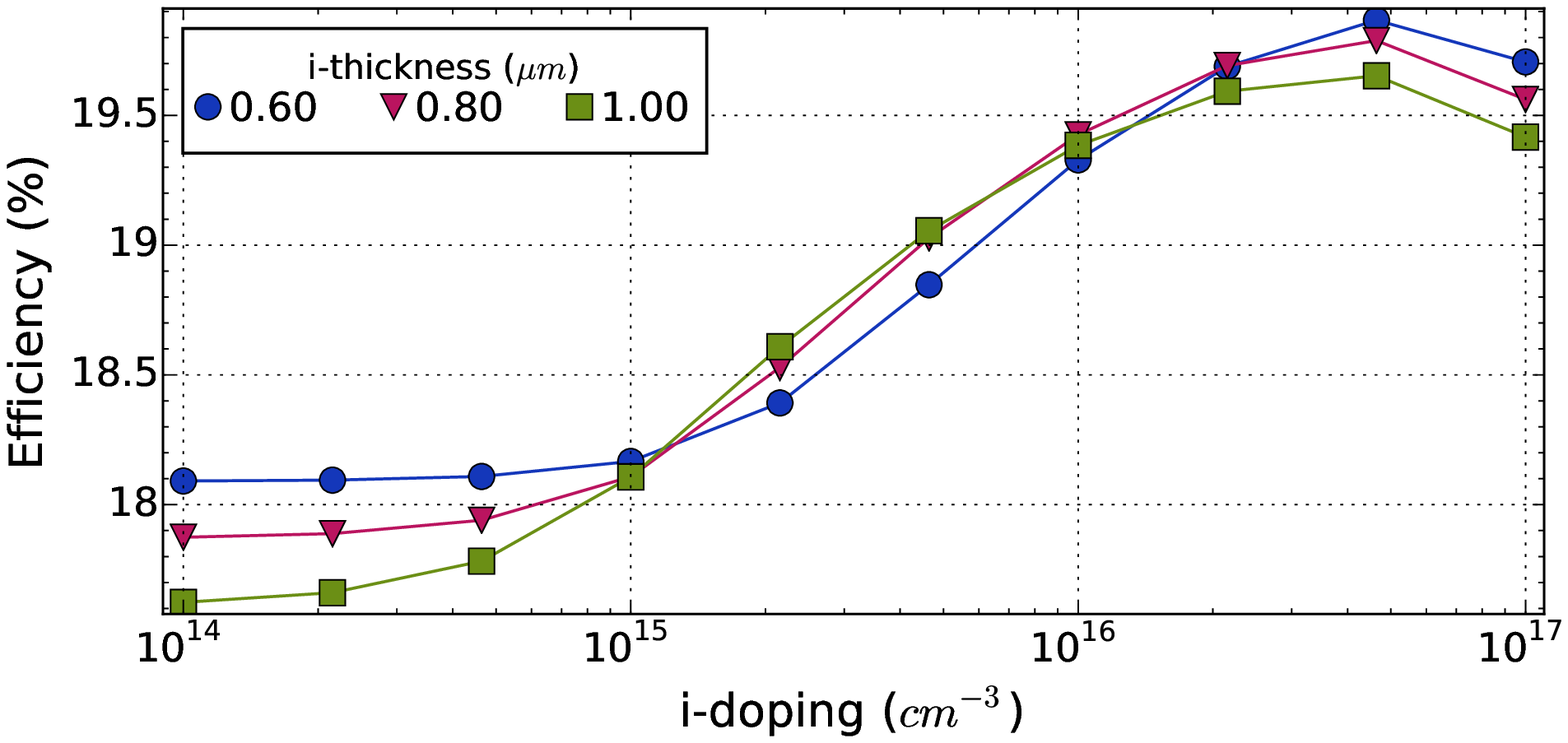}
	\label{min_eff_dop}
}
\caption{InGaN MIN solar cell efficiency for the optimal parameters given in table \ref{tab2_Optimum}, varying only the i-layer parameters. Only part of the points is plotted for clarity's sake.}
\label{Curves_MIN}
\end{figure}

As demonstrated in the previous section, the optimal $PN$ and $PIN$ solar cell efficiencies were obtained for p-layer thicknesses much lower than the light penetration depth and for a relatively high p-doping. The $Schottky$ solar cell, obtained by replacing the p-layer in the $PN$ structure by a relatively high workfunction metal, was previously demonstrated as a reliable alternative to the InGaN $PN$ solar cell \cite{ould_saad_hamady2016numerical}. Similarly, replacing the p-layer in the $PIN$ solar cell by a rectifying metal/InGaN contact leads to the new $MIN$ (Metal-IN) structure. Figure \ref{SCH_MIN} schematically displays these $Schottky$ and $MIN$ structures.

The $Schottky$ and $MIN$ solar cells were optimized with respect to their most important parameters: $\Li$ and $\Ln$, the thicknesses of the $I$ and $N$ layers respectively and where applicable, $\Ni$ and $\Nd$, the doping levels of the $I$ and $N$ layers respectively and where applicable, the Indium composition $\X$ and the metal workfunction $\WF$. The optimization was conducted in the same way as in the previous section. As was also done in the previous section, the optimum efficiency is reported in table \ref{tab2_Optimum}, along with the associated photovoltaic parameters as well as the corresponding parameters and their tolerance range, with the same definition as in the previous section.

For the Schottky structure, we found a maximum cell efficiency of $18.2\%$ for the following optimal parameter values:
$\Ln=0.86 \mathrm{\mu m}$, $\Nd=6.5\times10^{16} \mathrm{cm^{-3}}$, $\X=0.56$ and $\WF= 6.30 \mathrm{eV}$. The corresponding open-circuit voltage is $V_{OC}=0.863 \mathrm{V}$ with a short-circuit current of $J_{SC}= \mathrm{26.80 mA/cm^{2}}$ and a fill factor of $FF=78.82\%$.

For the $MIN$ structure, the maximum cell efficiency is $19.8\%$ for the following parameters values:
$\Li=0.61 \mathrm{\mu m}$, $\Ln=0.83 \mathrm{\mu m}$, $\Ni=6.1\times10^{16} \mathrm{cm^{-3}}$, $\Nd=3.6\times10^{17} \mathrm{cm^{-3}}$, $\X=0.60$ and $\WF=6.30 \mathrm{eV}$.  The corresponding open-circuit voltage is $V_{OC}=0.835 \mathrm{V}$ with a short-circuit current of $J_{SC}=\mathrm{30.29 mA/cm^{2}}$ and a fill factor of $FF=78.39\%$.

Figure \ref{sbsc_jv} shows the current-voltage characteristics of the $Schottky$ and $MIN$ solar cells. We observe that the $Schottky$ structure has a lower $J_{SC}$ and a higher $V_{OC}$ compared to the $MIN$ structure. This is due to the different optimal Indium composition: the $Schottky$ structure has an optimal Indium composition of $56\%$, that is lower than the optimal value for the $MIN$ structure ($60\%$). $V_{OC}$ increases as the Indium concentration decreases, owing to the widening of the bandgap. Simultaneously, $J_{SC}$ decreases as the direct consequence of a lower solar light absorption. 

Figure \ref {min_eff_thick} shows the variation of the PV efficiency as a function of the i-layer thickness (i-thickness), whereas figure \ref{min_eff_dop} shows it as a function of i-doping, for different i-thicknesses. The optimal i-thickness value, as shown in figure \ref{min_eff_thick}, is about $0.60 \mathrm{\mu m}$ as a consequence of the trade-off between the solar light absorption, increasing with the thickness, and the diffusion length that need to remain relatively higher than the layer thickness.
The same figure \ref{min_eff_dop} shows that the optimal i-doping value is $6.1\times10^{16}cm^{-3}$, corresponding to the optimal Space Charge Region (SCR) in the device. 

In addition to its main advantage of being p-layer free, the $MIN$ structure has another decisive advantage over the $PN$ and even the Schottky structures: the wider tolerance ranges of its optimal parameters, as Table \ref{tab2_Optimum} shows. This is due to the additional degree of freedom obtained with the i-layer.
Indeed, for the Schottky structure, the tolerance range of the n-layer thickness is $[0.53 - 1.00] \mu m$, while for the $MIN$ structure, it is $[0.10 - 1.00] \mu m$. This gives the $MIN$ structure a wider n-layer manufacturing tolerance than the Schottky structure. This tolerance range is important when actual device fabrication is considered.

For the n-doping, the $Schottky$ structure has a tolerance range of $[1.0\times10^{16}-3.0\times10^{17}] cm^{-3}$, while, for the $MIN$ structure, the tolerance range is $[1.8\times10^{16}-1.0\times10^{19}] cm^{-3}$. This allows to design heavily dopped n-layer to elaborate low resistance ohmic contact on InGaN, one of the major challenges in the III-Nitride solar cell processing \cite{bhuiyan2012ingan}, and without noticeably impacting the PV performances.

\section{MIN structure with actual experimental InGaN composition, thickness and metal workfunction}
\label{actualInGaN}

The above presented optimisation work lead to an optimal InGaN composition of $\X=0.60$ which is not yet experimentally achieved with sufficient material quality, although some very recent papers suggest that these compositions are in the process of being accessible \cite{fabien2015large, krishna2016epitaxial, dinh2016movpe, yang2016growth}. In this section, we propose to use one actual recent Indium composition obtained by Fabien \emph{et al.} \cite{fabien2015large}, that is $\X=0.22$ for large-area solar cells, and to evaluate the maximum efficiency that it allows.

Furthermore, a thickness constraint is linked to a composition constraint. We therefore limited the reachable thickness to $0.4\mathrm{\mu m}$.

\begin{table}
	\setlength{\tabcolsep}{3pt}
	\footnotesize
	\centering
	\renewcommand{\arraystretch}{1.0}
\begin{tabular}{|l|c|c|}
\hline
Indium Composition $x$ & Defect energy ($eV$) & Concentration ($\mathrm{cm^{-3}}$) \\ \hline
$0.09$ & $3.05$ & $2.7\times10^{16}$ \\ \hline
$0.13$ & $2.76$ & $8.5\times10^{15}$ \\ \hline
$0.20$ & $2.50$ & $6.1\times10^{16}$ \\ \hline
\end{tabular}
\caption{The dominating deep-level defect parameters in InGaN as experimentally measured and reported in \cite{nakano2014electrical, lozac2012study} for the $x = 0.09$ Indium composition, in \cite{armstrong2012quantitative} for $x = 0.13$ and in \cite{gur2011detailed} for $x = 0.20$. The defect energy is measured relatively to the conduction band edge.}
\label{tab_Defect}
\end{table}

Even though, the actually grown layers can have a high density of defects\cite{yamamoto_metal-organic_2013}. To take it into account, we introduced, on the one hand, valence and conduction band Urbach tails in the simulation, with an energy of $0.125 \mathrm{eV}$ as experimentally obtained in \cite{valdueza2014high}, and, on the other hand, a Gaussian distribution of defects in the bandgap. We used defects that were experimentally studied in the literature using the well known Deep Level (Transient \& Optical) Spectroscopy (DLTS and DLOS), the Steady-State PhotoCapacitance (SSPC) and the Lighted Capacitance-Voltage (LCV) techniques \cite{nakano2014electrical, lozac2012study, armstrong2012quantitative, gur2011detailed} as summarized in table \ref{tab_Defect}. The capture cross section that we chose to include in the simulation is the highest experimental value reported in \cite{gur2011detailed}. 

\begin{table*}
	\setlength{\tabcolsep}{3pt}
	\footnotesize
	\centering
	\renewcommand{\arraystretch}{1.0}

	\resizebox{\textwidth}{!}{\begin{tabular}{|>{\bfseries}l|c|c|c|c|c|c|c|}

	\cline{2-8}\multicolumn{1}{c|}{} &  &  &  &  &  &  & $\bm{\eta(\%)}$ \\

	\multicolumn{1}{c|}{} & $\bm{\Li(\mu m)}$ & $\bm{\Ln(\mu m)}$ & $\bm{\Ni(cm^{-3})}$ & $\bm{\Nd(cm^{-3})}$ & $\bm{\WF(eV)}$ & $\bm{x}$ &  $\bm{V_{OC}(V)}$ \\

	\multicolumn{1}{c|}{} &  &  &  &  &  &  & $\bm{J_{SC}(mA/cm^2)}$ \\

	\multicolumn{1}{c|}{} &  &  &  &  &  &  & $\bm{FF(\%)}$ \\

	\hline
		
		Range & $\bm{[0.10 - 0.40]}$ & $\bm{[0.10 - 0.40]}$ & $\bm{[1.0\times10^{14} - 1.0\times10^{17}]}$ & $\bm{[1.0\times10^{16} - 1.0\times10^{19}]}$ & $\bm{[5.50 - 6.30]}$ & \shortstack[c]{\\ $\bm{[0 - 0.22]}$} & \\

	\hline
	
	       &  &  &  &  &  &  & $\bm{7.25}$ \\

      MIN & $\bm{0.40}$ & $\bm{0.40}$ & $\bm{6.27\times10^{15}}$ & $\bm{7.46\times10^{16}}$ & $\bm{6.30}$ & $\bm{0.22}$ & $\bm{1.438}$ \\

	       & $\bm{[0.18 - 0.40]}$ & $\bm{[0.29 - 0.40]}$ & $\bm{[4.6\times10^{14} - 2.2\times10^{16}]}$ & $\bm{[2.0\times10^{16} - 1.0\times10^{18}]}$ & $\bm{[6.20 - 6.30]}$ & \shortstack[c]{\\ $\bm{[0.19 - 0.22]}$} & $\bm{5.92}$ \\

	       &  &  &  &  &  &  & $\bm{85.2}$ \\

	\hline

	\end{tabular}}

	\caption{Optimum efficiency $\eta$ obtained for the $MIN$ solar cell with a recently published actual experimental $x=0.22$ Indium composition\cite{fabien2015large} and layer thicknesses limited to $0.4\micron$. The associated open-circuit voltage $V_{OC}$, short-circuit current $J_{SC}$ and Fill Factor $FF$, along with the corresponding physical and material parameters, are equally shown. For each parameter, a range and a tolerance range are given. The range, within which the optimum value of a given parameter is sought, is on the second line of the table. The tolerance range is given just below each parameter optimal value. It corresponds to the set of values of that parameter for which the efficiency $\eta$ remains above $90\%$ of its maximum, the other parameters being kept at their optimum values.}
	\label{tab_exp_optimum}

\end{table*}

The optimization process was then run within these constraints and yield the optimal parameters summarized in table \ref{tab_exp_optimum}. As could be expected the layer thicknesses as well as the Indium concentration were found at their maximum authorized value, $0.4\micron$ and $x=0.22$ respectively, yielding a 7.25\% maximum efficiency. However, the computed tolerances deserve attention, since they are higher than one fourth, or even one half, of the optimal values, as far as the thicknesses are concerned.

As this was carried out without defects included, we then evaluated the MIN cell efficiency while varying the total density of states from $1.0\times10^{13} \mathrm{cm^{-3}}$ to $1.0\times10^{17} \mathrm{cm^{-3}}$. This latter density is even higher than the dominating defects concentration reported in \cite{nakano2014electrical, lozac2012study, armstrong2012quantitative, gur2011detailed}.

\begin{figure}
\includegraphics[width=\linewidth]{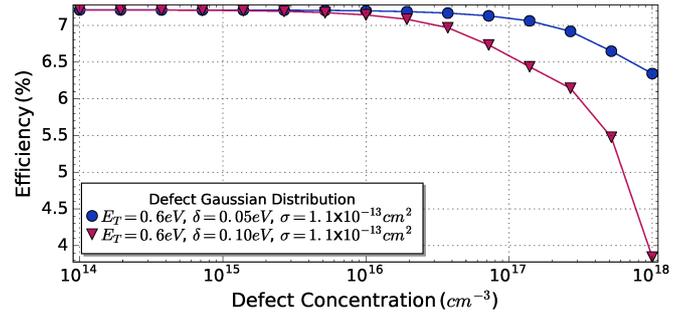}
\caption{The photovoltaic efficiency of the InGaN MIN solar cell with the actual experimental Indium composition, with defect density for two Gaussian distributions. The cell parameters are fixed to their optimal values shown in table \ref{tab_exp_optimum}.}
\label{wahab_fig7}
\end{figure}

Figure \ref{wahab_fig7} shows the MIN solar cell photovoltaic efficiency, with the actual experimental Indium composition, with respect to the defect concentration for two decay energy $\delta$ values of $0.05 \mathrm{eV}$ and $0.10 \mathrm{eV}$. The efficiency remains close to its maximum value as long as the defect concentration is smaller than the i-layer doping concentration ($6.1\times 10^{16}\mathrm{cm}^{-3}$). When the defect concentration becomes comparable to the optimal i-layer doping concentration, the solar cell efficiency decreases within a concentration range that depends on the distribution decay energy. This result means that the defects concentration must be kept lower but not necessarily much lower than the doping concentration. The demonstrated wide tolerance of the MIN structure can allow keeping as low as possible the negative impact of the defects on the overall solar cell efficiency,  by adjusting accordingly the InGaN doping. A compromise can therefore be found to limit the effect of the defects density that is relatively high in the presently elaborated InGaN layers.

\begin{table}
	\setlength{\tabcolsep}{3pt}
	\footnotesize
	\centering
	\renewcommand{\arraystretch}{1.0}
\begin{tabular}{|l|c|c|}
\hline
$W_f (eV)$ & $\eta$ for $\X=0.60$ (\%) & $\eta$ for $\X=0.22$ (\%) \\ \hline
$5.65$ & $6.34$ & $3.70$ \\ \hline
$5.93$ & $13.58$ & $5.22$ \\ \hline
$6.10$ & $17.84$ & $6.15$ \\ \hline
$6.30$ & $19.80$ & $7.25$ \\ \hline

\end{tabular}
\caption{Optimum efficiency $\eta$ obtained for a $MIN$ solar cell with various usual metal work functions $W_f$, lower than the optimal $6.30\mathrm{eV}$ yield by the optimization process; and for the optimal $\X=0.60$ Indium composition alongside the $\X=0.22$ composition published in \cite{fabien2015large}.}
\label{tab_Workfunction}
\end{table}

Finally, one may spot that the optimal $6.30\mathrm{eV}$ work function  obtained in table \ref*{tab_exp_optimum} seems to be relatively high when compared to the most reported values in the literature for Platinum (Pt), which is the ideal candidate for the practical realization of the MIN solar cell. However, a closer look at the reported values reveals a large dispersion in the Platinum work function measurements, from 5.65eV in \cite{rotermund1990investigation} to 6.35eV in the historical works by Lee Alvin Dubridge from Caltech (see \emph{e.g.} \cite{dubridge1928photoelectric}), through 5.93eV \cite{book2008work} and 6.10eV \cite{derry1989work}. To take these discrepancies into account, as well the possibility to use lower work function metals for the practical realization of the MIN solar cell, we have evaluated the foreseen efficiency for a set of possible work functions for both the optimal $x=0.60$ Indium composition and the $x=0.22$ composition reported in \cite{fabien2015large}. The results are summarized in table \ref{tab_Workfunction}.

% -----------------------------------------------------------------------------------------

\section{Conclusion}
\label{conclusion}

We investigated the photovoltaic performances of InGaN based $PN$, $PIN$ and $SBSC$ structures, using rigorous multivariate numerical optimization methods to simultaneously optimize the main physical and geometrical parameters of the solar cell structures.
We have found optimal photovoltaic efficiencies of $17.8\%$ and $19.0\%$ for the $PN$ and $PIN$ structures respectively. The optimization results led us to propose a new p-layer free $SBSC$ structure called $MIN$, the optimal efficiency of which is a higher $19.8\%$ for an Indium composition of a yet-to-reach $\X=0.60$, and as high as $7.25\%$ for a recent experimental Indium composition of $\X=0.22$ for a $0.4\micron$ thin layer that is not free of cristalline defects, the density of which we took into account. In addition, the $MIN$ structures has been shown to allow wider tolerance ranges on its physical and geometrical parameters, which allows to enhance its practical feasibility and reliability. The wider tolerance ranges of the new $MIN$ structure allow, for example, when compared to the previously studied Schottky structure, the easier realization of low resistance ohmic contacts, solely by raising the n-doping, as it was shown not to impair the efficiency.

% -----------------------------------------------------------------------------------------

%\section*{References}

\bibliographystyle{elsarticle-num} 
\bibliography{wahab_ref}

\end{document}